\documentclass[aps,prd,showpacs,preprintnumbers,nofootinbib,floatfix,onecolumn]{revtex4}

\usepackage{bm}
\usepackage{latexsym}
\usepackage{dcolumn}
\usepackage{amsfonts,amssymb}
\usepackage{graphicx,epsfig}
\usepackage{psfrag}

\def\beq{\begin{equation}}
\def\eeq{\end{equation}}
\def\br{\begin{eqnarray}}
\def\er{\end{eqnarray}}
\def\benu{\begin{enumerate}}
\def\eenu{\end{enumerate}}
\def\nn{\nonumber} 
\def\pa{{\partial}}
\def\l{\left}
\def\r{\right}    
\def\Hbar{\mathcal H}

\begin{document}
  

\title{Initial state of matter fields and trans-Planckian 
physics:~Can CMB\\ 
observations disentangle the two?}
\author{L.~Sriramkumar}
\email[]{E-mail: sriram@mri.ernet.in}
\affiliation{Harish-Chandra Research Institute, Chhatnag Road,
Jhunsi, Allahabad 211 019, India.}
\author{T.~Padmanabhan}
\email[]{E-mail: paddy@iucaa.ernet.in}
\affiliation{IUCAA, Pune University Campus, Post Bag 4,
Ganeshkhind, Pune 411 007, India.}

\date{\today}


\begin{abstract} 
The standard, scale-invariant, inflationary perturbation spectrum will 
be modified if the effects of trans-Planckian physics are incorporated 
into the dynamics of the matter field in a phenomenological manner, say, 
by the modification of the dispersion relation. 
The spectrum also changes if we retain the standard dynamics but modify 
the initial quantum state of the matter field.
We show that, given {\it any}\/ spectrum of perturbations, it is possible 
to choose a class of initial quantum states which can exactly reproduce 
this spectrum with the standard dynamics. 
We provide an explicit construction of the quantum state which will 
produce the given spectrum.  
We find that the various modified spectra that have been recently 
obtained from `trans-Planckian considerations' can be constructed 
from suitable squeezed states above the Bunch-Davies vacuum in the 
standard theory.  
Hence, the CMB observations can, at most, be useful in determining 
the initial state of the matter field in the standard theory, but 
it can {\it  not}\/ help us to discriminate between the various Planck 
scale models of matter fields.  
We study the problem in the Schrodinger picture, clarify various 
conceptual issues and determine the criterion for negligible back 
reaction due to modified initial conditions.
\end{abstract}
\maketitle


\section{The tangled web}\label{sec:intro}

The inflationary scenario~\cite{txts,ll99,infltnmdls} is at present 
the most attractive paradigm for generating the initial small scale 
inhomogeneities~\cite{dp,ps}. 
These perturbations leave their imprints as anisotropies in the 
Cosmic Microwave Background (CMB)~\cite{WMAP} and later evolve, 
due to gravitational instability, into the large-scale structures 
that we see around us today. 
While there is no natural particle physics candidate for generating 
the inflationary phase, a single (or a few) `inflaton' field(s), with 
fine-tuned designer couplings, is (are) often introduced in order to 
reproduce the observed magnitude and shape of the perturbation spectrum.

In many of the models of inflation~\cite{infltnmdls}, the period of 
acceleration lasts sufficiently long so that length scales that are 
of cosmological interest today would have emerged from sub-Planckian 
length scales at the beginning of inflation.
This suggests that physics at the very high energy scales can, in 
principle, modify the primordial perturbation spectrum~\cite{tplp} 
and these modifications can---in turn---leave their signatures on 
the CMB~\cite{mdr1}.
This has led to a considerable interest in understanding the 
effects of Planck scale physics on the inflationary perturbation 
spectrum~\cite{tplp}--\cite{mb03} and the 
CMB~\cite{minimalcmb,nccmb,lqgcmb}.

Metric fluctuations during the inflationary epoch can be 
modeled by a quantized, massless and minimally coupled 
scalar field~\cite{dp,ps,mfb92}. 
In the absence of a workable quantum theory of gravity, the Planck 
scale effects on the perturbation spectrum have to be studied by 
phenomenologically modifying the dynamics of the scalar field to 
take into account the quantum gravitational effects (for an early 
attempt in this direction, see Ref.~\cite{tplp}).
The high energy models of the quantum scalar field that have been 
popular in the literature either introduce new features in the 
dispersion relation~\cite{tpzero,mdr1,mdr2,adbtcty,br} or modify 
the standard uncertainty principle~\cite{othr,gup} or assume that
the spacetime coordinates are non-commutative~\cite{nc}. 
(For other approaches, see Refs.~\cite{minimal,gnrl,burgess,ss04} 
and, for a recent review of many of these approaches, see 
Ref.~\cite{mb03}).
Some of these models have been utilized to evaluate not only the 
Planckian corrections to the standard, scale-invariant perturbation 
spectrum, but also the resulting signatures on the 
CMB~\cite{minimalcmb,nccmb,lqgcmb}.  
This suggests the possibility that sufficiently accurate measurements 
of the CMB anisotropies can help us understand physics  beyond the 
Planck scale.

There is, however, one serious difficulty with this approach, which 
we shall now briefly describe.

We begin by noticing that, in a Friedmann universe, each mode $q_{\bf k}$ 
of the scalar field, labeled by the wave vector~${\bf k}$, evolves as an 
independent oscillator with time dependent parameters that are related to 
the expansion factor $a(t)$. 
Given the quantum state $\psi_{\bf k}[q_{\bf k}, t_i({\bf k})]$ for 
the mode $q_{\bf k}$ at a time $t_i({\bf k})$, one can obtain the 
state at a later time $t$ by
\beq
\psi_{\bf k}(q_{\bf k}, t)
=\int d{\bar q}_{\bf k}\, K[q_{\bf k},t;\; 
{\bar q}_{\bf k}, t_i({\bf k})]\; 
\psi_{\bf k}[{\bar q}_{\bf k}, t_i({\bf k})],
\label{eq:krnl}
\eeq
where $K(q_{\bf k},t;\; {\bar q}_{\bf k},t_i)$ is the (path integral) 
kernel for an oscillator with time dependent parameters, which can be 
written down in terms of the classical solution (see, for e.g., 
Ref.~\cite{dr94}). 
To keep our discussion general, we have allowed for the 
possibility that the initial quantum state of each of the oscillators 
is specified at different times so that $t_i$ can depend on ${\bf k}$. 
For instance, one may choose to specify the quantum state of each 
oscillator when the proper wavelength of that mode is equal to 
the Planck length~\cite{minimal,state}. 
Of course, when the initial state of all the modes are specified at 
a given time, $t_i({\bf k})=t_i$ will be independent of ${\bf k}$. 
The {\it dynamics}\/ of the system is completely specified by the 
kernel~$K$.

It is obvious that the mathematics of a free scalar field in a 
Friedmann universe is trivial and is no more complicated than 
that of an oscillator with time dependent parameters. 
The perennial interest in this problem (allowing so many e-prints 
to be written!) arises from two conceptual issues: 
\begin{enumerate}
\item
To make any predictions, we need to know $\psi_{\bf k}(q_{\bf k}, t)$, 
which, in turn, requires knowing $\psi_{\bf k}[{\bar q}_{\bf k}, 
t_i({\bf k})]$.
We have absolutely no idea what to use for $\psi_{\bf k}[{\bar q}_{\bf k}, 
t_i({\bf k})]$ and so different choices (usually called `vacuum states', 
which is only a manner of speaking) can be investigated.
\item
There are  infinite number of such oscillators, leading to the 
standard, unresolved, issues of regularization in quantum field 
theory. 
\end{enumerate}
To make any progress, we need to {\it make an assumption}\/ regarding 
$\psi_{\bf k}[{\bar q}_{\bf k}, t_i({\bf k})]$ and our results are 
only as valid as this assumption. 

If we now further modify the dynamics (due to a phenomenological 
input regarding trans-Planckian physics), we will be changing the 
form of the kernel $K$. 
But, since we can only observe the integrated effect of $K$ and 
$\psi_{\bf k}[{\bar q}_{\bf k}, t_i({\bf k})]$, the observations 
can tell us something about the $K$ (and trans-Planckian physics) 
only if we assume something about $\psi_{\bf k}[{\bar q}_{\bf k}, 
t_i({\bf k})]$. 
The usual assumption is to consider the initial quantum state to be 
the Bunch-Davies vacuum~\cite{bd78}, but it is only an assumption.
The crucial question is whether the effect of trans-Planckian physics 
can be mimicked by a different choice of the initial state other than 
the Bunch-Davies vacuum. 

We shall show that {\it any}\/ (modified) spectrum of fluctuations 
can be obtained from a suitably chosen initial state 
$\psi_{\bf k}[{\bar q}_{\bf k}, t_i({\bf k})]$, which will prove to 
be a squeezed state above the Bunch-Davies vacuum in the standard 
theory.
We shall provide an explicit construction of the state for any given 
spectrum of perturbations that is observed.
So, if some specific deviation from the standard scale invariant 
spectrum is seen in the CMB, a conservative interpretation will 
be to attribute it to a deviation from the standard initial state 
of the theory. 
Unless this possibility is ruled out, one cannot claim that the 
observation supports, say, a particular model of trans-Planckian 
phenomenology.
Motivated by this result, we argue that the CMB can at most help us 
identify the quantum state of the scalar field in the standard theory, 
but it can {\it not}\/ aid us in discriminating between the various 
Planck scale models of matter fields.

The remainder of this paper is organized as follows.
In Section~\ref{sec:gf}, we set up the formalism and study the evolution 
of a Gaussian quantum state in a Friedmann universe. 
We apply this formalism to power law inflation in Section~\ref{sec:plei}.
In Section~\ref{sec:mtpe}, we show that any modified spectrum can be 
reproduced from a suitable squeezed state above the Bunch-Davies vacuum 
in the standard theory.
We explicitly discuss four modified spectra that have recently been 
considered in the literature.
In Section~\ref{sec:msbr}, we evaluate the energy density in these
excited states and examine whether these modified spectra also lead 
to a large back reaction on the inflating background.
Finally, in Section~\ref{sec:dscssn}, we conclude with a discussion
on the wider implications of our analysis.

Our conventions and notations are as follows. 
We shall set $\hbar=c=1$ and the metric signature we shall adopt is 
$(+, -,-, -)$.

\section{Evolution of the quantum state:~General formalism}\label{sec:gf}

Consider a flat Friedmann universe described by the line-element
\beq
ds^2=dt^2-a^{2}(t)\, d{\bf x}^2
= a^{2}(\eta)\l(d\eta^{2} - d{\bf x}^2\r),\label{eq:frwle}
\eeq 
where $t$ is the cosmic time, $a(t)$ is the scale factor and $\eta$ 
denotes the conformal time with $d\eta= dt/a(t)$. 
The scalar as well as the tensor perturbations during the inflationary 
epoch can be modeled by a massless and minimally coupled, real scalar 
field, say, $\Phi$, governed by the 
action~\cite{txts,ll99,dp,ps,mfb92,paddy03}
\beq
S[\Phi] =\frac{1}{2}\int d^4x\, \sqrt{-g}\; 
\pa_{\mu}\Phi\, \pa^{\mu}\Phi.\label{eq:actn}
\eeq
The homogeneity and isotropy of the Friedmann metric~(\ref{eq:frwle}) 
allows us to decompose the scalar field $\Phi$ as
\beq
\Phi({x})
= \int\frac{d^{3}{\bf k}}{(2\pi)^{3/2}}\, 
\bar q_{\bf k}(\eta)\, e^{i{\bf k}\cdot{\bf x}}
\label{eq:defPhi}
\eeq  
The $\bar q_{\bf k}$ is complex and for each ${\bf k}$ gives two degrees of freedom
in the real and imaginary parts of $\bar q_{\bf k}=A_{\bf k}+iB_{\bf k}$. But
since $\Phi$ is a real scalar field, we can relate the variables for $ {\bf k}$
to that for $-{\bf k}$ by $A_{\bf k}^{\ast}=A_{\bf -k}, B_{\bf k}^{\ast}=-B_{\bf -k}$ 
and only half the modes are independent degrees of freedom.
Therefore, we can work with new set of  the real  modes $q_{\bf k}$ for all
values of ${\bf k}$ with suitable redefinition, say, by taking $q_{\bf k}=A_{\bf k}$ for one half
of ${\bf k}$ vectors and $q_{-\bf k}=B_{\bf k}$ for the other half. In
terms of real variables $q_{\bf k}$, the action~(\ref{eq:actn}) can 
be expressed as follows:
\beq
S[\Phi]=\frac{1}{2}\int d^{3}{\bf k}\int d\eta\; a^2\,
\l(q_{\bf k}'^{2} -k^2\, q_{\bf k}^2\r),
\label{eq:actnq}
\eeq
where the primes denote differentiation with respect to the conformal 
time $\eta$ and  $k=\vert {\bf k}\vert$.
The action~(\ref{eq:actnq}) describes a collection of independent
oscillators with time dependent mass~$a^2$ and frequencies $k$.
(It is sometimes useful to keep track of the real and imaginary parts 
of the $\bar q_{\bf k}$ separately.  
In our case, this is unnecessary and we can define our system in terms 
of real $q_{\bf k}$. 
Our approach is completely equivalent to the conventional one.   
Also, note that, in the literature, one usually finds the Fourier 
decomposition in Eq.~(\ref{eq:defPhi}) expressed in terms of 
$(q_{\bf k}/a)$ rather than with just $q_{\bf k}$. 
Such a decomposition will lead to oscillators that have a unit mass,
but a time-dependent frequency, say, $\omega^2_{\bf k}$, which can become
negative at super Hubble scales. 
In our description---which is, again, equivalent to the conventional
one---the mass varies as $a^2$, but the frequency is constant.)
 
In the Schrodinger picture, the scalar field $\Phi$ can be quantized 
by quantizing each independent oscillator~$q_{\bf k}$.
The Hamiltonian corresponding to the ${\bf k}$-th oscillator is given 
by
\beq
{\sf H}_{\bf k}=\frac{p_{\bf k}^2}{2\, a^2}
+\frac{1}{2}\,a^2 k^{2} q_{\bf k}^2,
\eeq
where $p_{\bf k}$ is the momentum conjugate to the 
coordinate $q_{\bf k}$.
Therefore, each of the oscillators satisfy the Schrodinger equation
\beq
i\, \frac{\pa \psi_{\bf k}}{\pa \eta}
=-\frac{1}{2\, a^2}\, \frac{\pa^2\psi_{\bf k}}{\pa q_{\bf k}^2}
+\frac{1}{2}\, a^{2} k^{2} q_{\bf k}^{2}\, \psi_{\bf k}\label{eq:seq}
\eeq
and the complete quantum state of the field is described by a wave 
function that is a product of $\psi_{\bf k}$ for all~${\bf k}$.
Equivalently, the time evolution of the wave function can be 
described by Eq.~(\ref{eq:krnl}) in terms of the kernel 
$K\l[q_{\bf k},t;\; {\bar q}_{\bf k},t_i({\bf k})\r]$. 

As we do not expect a large scale, spatially inhomogeneous classical 
scalar field to be present in the universe, it is conventional to 
assume that the expectation value $\langle \psi_{\bf k}\vert 
{\hat q}_{\bf k} \vert \psi_{\bf k} \rangle$ vanishes in the quantum 
state of the field for ${\bf k} \neq 0$. 
(Since $\langle \psi_{\bf k} \vert {\hat q}_{\bf k} \vert \psi_{\bf k}
\rangle$ satisfies the classical equations of motion, this condition 
can be satisfied at all times if suitable initial conditions are imposed 
at an early epoch.)
When the mean value vanishes, the power spectrum as well as the 
statistical properties of the perturbations are completely 
characterized by the two-point functions of the quantum field.
Therefore, the power spectrum of the perturbations per logarithmic 
interval, viz. $\left[k^3\, {\cal P}_{\Phi}(k)\right]$, is given by 
(see, for instance, Ref.~\cite{paddy03})
\beq
k^3\; {\cal P}_{\Phi}(k)
=\frac{k^3}{2\pi^2}\, 
\int\limits_{-\infty}^{\infty} dq_{\bf k}\, 
\vert \psi_{\bf k}\l(q_{\bf k},\eta\r)\vert^2\, 
q_{\bf k}^{2}.\label{eq:psdfntn}
\eeq
Though this result is fairly well established  in literature, there 
are a couple of subtleties we would like to mention.

The quantity on the right hand side of the above equation depends on 
the time $\eta$ and one needs to settle at what epoch it has to be 
evaluated.
In classical perturbation theory, one evaluates the perturbation at 
Hubble exit, i.e. when the physical wavelength $(k/a)^{-1}$ of the 
mode corresponding to the wavenumber ${\bf k}$ {\it is comparable to}\/ 
the Hubble radius $H^{-1}$, where $H=(a'/a^2)$.  
In other words, the spectrum is to be evaluated when, say, $(k/a)
= (z\,H)$, where $z$ is a number of the order of unity.
This is a thumb rule which accounts for the differences in the 
evolutionary history of the mode when its proper wavelength is 
smaller than the Hubble radius as compared to the situation when 
it is larger than the Hubble radius. 
There is no simple way of deciding whether one should evaluate the 
expression when $z= 1$ or when, say, $z= \l(2\pi\r)$.  
In the literature, one also finds the perturbation spectrum evaluated 
at super Hubble scales (i.e. when $(k/aH)\to 0$) which corresponds
to the limit $z\to 0$.  
Even for the simplest case of exponential inflation with the Bunch-Davies 
vacuum for the initial state, these results differ by a numerical factor.
In this case, we will find that, $[k^3\; {\cal P}_{\Phi}(k)]=\l[C(z)\, 
\l(H^2/2 \pi^2\r)\r]$, where $C(z)=[(1+z^2)/2]$ (see Eq.~(\ref{eq:psefr}) 
below). 
It is sometimes claimed erroneously in the literature, that evaluating 
the spectrum as $z\to 0$ leads to the same result as evaluating it at 
$z=1$. 
(See, for instance, Ref.~\cite{ll99}, pp.~182--183. 
Notice that, in this reference, Eq.~(7.87) is wrong by a factor $2$ and 
the claim after Eq.~(7.98) that Eqs.~(7.98) and~(7.87) are ``in agreement" 
is incorrect. 
This can be trivially verified from equations~(7.96) and (7.98).)
In this particular case (i.e. exponential inflation with the Bunch-Davies 
vacuum for the initial state), this discrepancy is not of great importance 
since it only changes the amplitude by a numerical factor, since
$C(1)=1$, $C(2\pi)=\l[\l(1+4\pi^2\r)/2\r]$ and $C(0)=(1/2)$.
But when one considers the power spectrum in a more general context, these 
choices will lead to a more complicated difference, as we shall see. 
Of course, if one first approximates the wave function $\psi_{\bf k}
\l(q_{\bf k}, \eta\r)$ with an assumption such as $z \to 0$ (when the 
expressions do simplify) {\it and then}\/ evaluate it at $z \simeq 1$, 
one is being inconsistent. 
This may sound rather elementary but we were surprised to find papers 
in the literature which do this.
Any result which crucially depends on one specific choice for $z$ in 
computing the power spectrum is suspect. 
In what follows, we shall usually assume that the expressions are evaluated 
for $z=1$ (i.e. when $(k/a) = H$) when the choice of $z$ is not of much 
consequence and will comment on the results which depend crucially on 
this choice.

Let us now consider the problem of determining the wave function 
$\psi_{\bf k}\l(q_{\bf k},\eta\r)$.
For a time dependent oscillator, there is no concept of a unique ground 
(`vacuum') state unless the parameters describing the oscillator go to 
a constant value asymptotically---which, in general, it does not, for 
the Friedmann universe.  
There is, however, a class of solutions to the time dependent harmonic
oscillator which are {\it form invariant}\/ in the sense that the $q_{\bf k}$ 
dependence of the wave function $\psi_{\bf k}(q_{\bf k}, \eta)$ is the same 
at all~$\eta$. 
One can show that the most general state having this property is an 
exponential of a quadratic function of $q_{\bf k}$ and such states have 
been extensively investigated in the past in different contexts (see, 
for instance,  Refs.~\cite{tpgswfn,ss}). 
When $\langle \psi_{\bf k}\vert {\hat q}_{\bf k} \vert\psi_{\bf k}\rangle 
=0$, the mean value of the Gaussian vanishes and the quantum state of the 
mode can be described by the wave function~\cite{paddy03,tpgswfn,ss}
\beq
\psi_{\bf k}\l(q_{\bf k},\eta\r)
= N_{k}(\eta)\, \exp-\l[R_{k}(\eta)\; q_{\bf k}^2 \r],
\label{eq:gswfn}
\eeq
where $N_{k}(\eta)$ and $R_{k}(\eta)$ are complex 
quantities. 
The normalization condition on the wave function then relates $N_{k}$ 
and $R_{k}$ as follows:
\beq
\vert N_{k}\vert^2=\l(\frac{R_{k}+R_{k}^\ast}{\pi}\r)^{1/2}.
\label{eq:NkRk}
\eeq 
Therefore, the only non trivial aspect of the quantum state is encoded 
in the time dependence of the function $R_{k}(\eta)$. 
It can be shown that (for details, see Appendix~\ref{app:eps}), 
if we introduce a function $\mu_k(\eta)$ through the relation 
$R_{k}=-\l(i\, a^2/2\r)(\mu_{k}'/ \mu_{k})$, then $\mu_k$ 
satisfies the differential equation
\beq
\mu_{k}''+2\, \frac{a'}{a}\, \mu_{k}'
+k^2 \mu_{k}=0\label{eq:deq}
\eeq
which is the same as the classical equation of motion satisfied by the 
oscillator variable $q_{\bf k}$.
We find that the perturbation spectrum~(\ref{eq:psdfntn}) corresponding 
to the wave function~(\ref{eq:gswfn}) can be expressed as (see 
Appendix~\ref{app:eps})
\beq
k^3\; {\cal P}_{\Phi}(k)
=\frac{k^3}{2\pi^2}\, 
\l(\frac{\vert \mu_{k}\vert^2}{W(k)}\r),\label{eq:ps}
\eeq
where $W(k)$ is a $k$-dependent constant determined by the Wronskian 
condition for $\mu_{k}$ [cf.~Eq.~(\ref{eq:wrnskn})].

Since the differential equation~(\ref{eq:deq}) has real coefficients, 
if $s_k$ is a solution, so is $s_k^*$ and the general solution is a 
linear superposition of the form, say, 
$\mu_k=\l[{\cal A}(k)\, s_{k} + {\cal B}(k)\, s_{k}^*\r]$. 
The quantum state $\psi_{\bf k}(q_{\bf k}, \eta)$, however, depends only 
on $R_{k}$ which is independent of the overall scaling of $\mu_k$. 
This feature translates into the power spectrum~(\ref{eq:ps}) as well; 
a global scaling of $\mu_{k}$ also changes the Wronskian~$W(k)$ leaving 
$\l[\vert \mu_k\vert^2/W(k)\r]$ invariant.  
Hence, we can ignore the overall scaling in $\mu_k$.
We shall set ${\cal A}(k)$ to unity and choose the Wronskian~$W(k)$ 
to be
\beq
W(k)=1-\vert {\cal B}(k)\vert^2.\label{eq:BW}
\eeq
Then, the power spectrum~(\ref{eq:ps}) reduces to
\beq
k^3\; {\cal P}_{\Phi}(k)
=\frac{k^3}{2\pi^2}
\l(\frac{\l[1+\vert {\cal B}(k)\vert^2\r] \vert s_{k}\vert^2 +2\, 
{\rm Re.}~[{\cal B}(k)\, {s_{k}^*}^2]}{1-\vert {\cal B}(k)\vert^2}\r).\;
\label{eq:psone}
\eeq
(In the Heisenberg picture, one usually introduces the Bogoliubov 
coefficients $\alpha(k)$ and $\beta(k)$.
These coefficients are related to ${\cal B}(k)$ through the relation: 
${\cal B}(k)=[\beta(k)/\alpha(k)]$.
In terms of $\alpha(k)$ and $\beta(k)$, the Wronskian 
condition~(\ref{eq:BW}) reduces to the standard normalization condition, 
viz.~$\l(\vert \alpha(k)\vert^2-\vert \beta(k)\vert^2\r)=1$.)
The dynamics imposes no restrictions on the form of ${\cal B}(k)$, 
which is related to the choice of the quantum state at any specified 
time. 
The spectrum evidently depends on ${\cal B}(k)$ and with a suitable 
choice for ${\cal B}(k)$ we can generate any, given, reasonable 
spectrum. 

\subsection{Squeezing and instantaneous particle content of the 
quantum state}

We shall see concrete examples of this result and its consequences 
for the case of power law inflation in the following sections.
But, before we do that, let us try and understand what the wave 
function~(\ref{eq:gswfn}) implies.
The wave function~(\ref{eq:gswfn}), in general, describes what is 
referred to in the literature as a squeezed state (see, for e.g., 
Refs.~\cite{ss}).
Squeezed states for a given mode $q_{\bf k}$ are described by two 
parameters, say, $r_{k}$ and $\varphi_{k}$ and the quantity $R_{k}$ 
can be related to these two parameters as follows \cite{ss}:
\beq
R_{k}=\left(\frac{k\, a^2}{2}\right)\,
\l(\frac{{\rm cosh}\,r_{k} 
+ e^{2 i\varphi_{k}}\, {\rm sinh}\,r_{k}}{{\rm cosh}\,r_{k}
- e^{2 i\varphi_{k}}\, {\rm sinh}\,r_{k}}\r).
\eeq
This, however, does not lead to any deeper insight in this particular 
case.

An alternative procedure, which is physically better motivated, is 
to compare  the quantum state~(\ref{eq:gswfn}) with the instantaneous 
ground state at any given time. 
We recall that the oscillators have the frequency $k$ and a time 
dependent mass $a^2(\eta)$.
The ground state of a harmonic oscillator with the frequency $k$ 
and mass $a^2(\eta)$ will depend on $q_{\bf k}$ as $\exp-[(k/2)\, 
a^2\, q_{\bf k}^2]$.  
Suppose we are given a specific quantum state determined by the 
function $R_k(\eta)$. 
We can attempt to solve the equation $R_k(\eta)=[k\, a^2(\eta)/2]$ 
for $\eta$ determining a function, say, ${\bar \eta}(k)$. 
If such a real solution can be obtained, then we can interpret our 
quantum state as the ground state for the mode $q_{\bf k}$ at the moment 
of time  ${\bar \eta}(k)$.
In general, however, this will {\it not}\/ be possible since $R_{k}(\eta)$ 
is complex. 
(One exception is in the adiabatic approximation, in which the mode 
$\mu_{k}$ is {\it chosen}\/ such that that $R_{k}\approx (k\, a^2/2)$ 
so that this condition is identically satisfied; as to be expected, 
the state evolves as an adiabatic vacuum in this case.) 

More generally, one can expand our quantum state in terms of the 
instantaneous states of the harmonic oscillator at any given time 
$\eta$, thereby identifying its particle content.
At any given time $\eta$, the {\it instantaneous}\/ ground and excited 
states of the oscillators $q_{\bf k}$ can be described by the wave 
functions
\beq
\phi_{n}\l(q_{\bf k},\eta\r)
=\l(2^{n}\, n!\r)^{-1/2}
\l(\frac{k\, a^2}{\pi}\r)^{1/4}\, 
H_{n}\l(\sqrt{k}\, a\, q_{\bf k}\r)\; \exp -\left(\frac{k}{2}\, 
a^2\, q_{\bf k}^2 + i E_{n}\, \eta\right),
\eeq
where $H_{n}$ are the Hermite polynomials, $E_{n}= [(n+ (1/2))\, k]$ 
and $n=0,1,2,\ldots$.
We can now decompose the actual wave function $\psi_{\bf k}\l(q_{\bf k},
\eta\r)$ in terms of the above instantaneous wave functions as follows:
\beq
\psi_{\bf k}\l(q_{\bf k},\eta\r)=\sum\limits_{n=0}^{\infty}
c_{n}(k, \eta)\, \phi_{n}\l(q_{\bf k},\eta\r)\, e^{i E_n \eta},
\eeq
where the decomposition amplitude $c_{n}$ is given by the integral 
\beq
c_{n}(k, \eta) =\int\limits_{\infty}^{\infty} dq_{\bf k}\; 
\psi_{\bf k}\l(q_{\bf k},\eta\r)\; \phi_{n}^{\ast}\l(q_{\bf k},\eta\r).
\eeq
On evaluating this integral, we find that the amplitude $c_{n}$
corresponding to the odd $n$'s vanish, while the amplitude for the
even $n$'s are given by
\beq
c_{(2n)}(k, \eta)=\Delta_{k}\, \l(\frac{\sqrt{(2n)!}}{2^{n}\, 
n!}\r)\, \Gamma_{k}^n,
\eeq
where
\beq
\Delta_{k}=N_{k}\, \l(\frac{k\, a^2}{\pi}\r)^{1/4}\,
\l(\frac{2\pi i\mu_{k}/a^2}{\mu_{k}'+ik\,\mu_{k}}\r)^{1/2}
\eeq
is a $n$-independent normalization and
\beq
\Gamma_{k}
=-\l(\frac{\mu_{k}'- ik\,\mu_{k}}{\mu_{k}'+ik\, \mu_{k}}\r).
\label{eq:Gamma}
\eeq
The probability $P_{k}(n)$ for our quantum state to be in the $(2n)$-th 
excited state of the instantaneous harmonic oscillator mode can be 
thought of as the probability for existence of $n$ pairs of particles 
at the time $\eta$. 
This is given by
\beq
P_{k}(n)\equiv \vert c_{(2n)}(k, \eta) \vert^2
=P_{k}(0)\, \left(\frac{(2n)!}{n!^2}\right)
\left(\frac{|\Gamma_{k}|^2}{4}\right)^{n}
\eeq
and the generating function ${\cal G}_{k}(\sigma)$ for this pair 
creation probability can be expressed in closed form as follows:
\beq
{\cal G}_{k}(\sigma) \equiv \sum_{n=0}^{\infty}P_{k}(n)\, \mu^{n}
=\frac{P_{k}(0)}{\left(1-\sigma\, |\Gamma_{k}|^2\right)^{1/2}}
=\l(\frac{1-|\Gamma_{k}|^2}{1-\sigma\, |\Gamma_{k}|^2}\r)^{1/2}.
\eeq
The last equation follows from explicitly calculating $P_{k}(0)$ 
or---more simply---by noticing that $P_{k}(n)$ is normalized and 
hence ${\cal G}_{k}(1)=1$. 
Given this generating function one can compute various moments of the 
created particles. 
In particular, the mean number of {\it particles},\/ which are present at 
the time $\eta$ (obtained by doubling the mean number of pairs) is given
by
\beq
\langle n_{k}\rangle 
=\frac{|\Gamma_{k}|^2}{1-|\Gamma_{k}|^2}.\label{eq:nk}
\eeq
These equations exhibit the time dependent particle content of our 
quantum state and it can be computed once the function $\mu_{k}(\eta)$
is specified. 
Though, $\langle n_{k}\rangle$ can not be interpreted as particles with 
respect to the in-vacuum at late times, it is related in a simple manner 
to the energy density. 
The expectation value of the Hamiltonian operator corresponding to the 
oscillator $q_{\bf k}$, say, ${\cal E}_{k}$, can be evaluated using the 
wave function~(\ref{eq:gswfn}). 
We obtain that
\beq
{\cal E}_{k}
=\l(\frac{a^2}{2\, W(k)}\r)\l(\vert \mu_{k}'\vert^2 
+k^2\, \vert \mu_{k}\vert^2\r)\label{eq:calEk}
\eeq
and, on using the expressions (\ref{eq:Gamma}), (\ref{eq:nk}) and 
(\ref{eq:calEk}), we find that  ${\cal E}_{k}$ and $\langle n_{k}
\rangle$ are related as follows:
\beq
{\cal E}_{k}=\l(\langle n_{k}\rangle +\frac{1}{2}\r)\, k.
\label{eq:calEknk}
\eeq
The energy density of the quantum scalar field is then given by
\beq
\rho=\frac{1}{2\pi^2 a^{4}}\,
\int\limits_{0}^{\infty} dk\; k^2\, 
{\cal E}_{k}=\frac{1}{2\pi^2 a^{4}}\,
\int\limits_{0}^{\infty} dk\; k^3\, \l(\langle n_{k}\rangle 
+\frac{1}{2}\r).\label{eq:ed}
\eeq
We will require these results while discussing the issue of back reaction
due to the modified initial conditions.

\subsection{Wigner function}

Another possible way of understanding the physical content of a quantum 
state, especially its classicality, is through the Wigner function (see, 
for e.g., Ref.~\cite{dchrnce}).  
Given a wave function $\psi_{\bf k}(q_{\bf k},\eta)$, the Wigner 
function ${\cal W}_{k}\l(q_{\bf k}, p_{\bf k}, \eta\r)$ is defined 
as~\cite{dchrnce}
\beq
{\cal W}_{k}\l(q_{\bf k}, p_{\bf k},\eta\r)
=\frac{1}{2\pi}\, \int\limits_{\infty}^{\infty}du_{\bf k}\;
\psi_{\bf k}^*\l[(q_{\bf k}+\frac{1}{2}u_{\bf k}),\eta\r]\; 
\psi_{\bf k}\l[(q_{\bf k}-\frac{1}{2}u_{\bf k}),\eta\r]\; 
e^{ip_{\bf k}u_{\bf k}}.
\eeq
The Wigner function corresponding to the Gaussian wave 
function~(\ref{eq:gswfn}) can be expressed as
\beq
{\cal W}_{k}\l(q_{\bf k}, p_{\bf k}, \eta\r)
=\frac{1}{\pi}\; 
\exp-\l[\frac{q_{\bf k}^2}{\sigma_{k}^2(\eta)}
+\sigma _{k}^2(\eta)\, \l(p_{\bf k}-
{\cal J}_{k}(\eta)\,  q_{\bf k}\r)^2\r],
\eeq
where $\sigma _{k}$ and ${\cal J}_{k}$ are given by
\beq
\sigma _{k}^2=\l(R_{k}+R_{k}^{\ast}\r)^{-1}
\qquad{\rm and}\qquad {\cal J}_{k}= i\,\l(R_{k}-R_{k}^{\ast}\r).
\eeq
On using the relations~(\ref{eq:Rk}) and (\ref{eq:wrnskn}), we find 
that the quantities $\sigma _{k}$ and ${\cal J}_{k}$ can be written
in terms of the function $\mu_{k}$ and the Wronskian $W(k)$ as follows:
\beq
\sigma _{k}^2=\l(\frac{2\, \vert \mu_{k}\vert^2}{W(k)}\r); 
\qquad
{\cal J}_{k}=\l(\frac{a^2}{2}\r)
\frac{d\, \ln |\mu_k|^2}{d\eta}.
\eeq
The quantum versus classical nature of the wave function depends on the evolutionary behaviour of $\sigma_k(\eta)$ and ${\cal J}_{k}(\eta)$. It is possible for evolution to lead to a Wigner function sharply peaked around some region in the phase space, starting from a Wigner function which is uncorrelated in phase space~\cite{wfds}.
For example, in the case of exponential inflation and the Bunch-Davies vacuum 
for the initial state which we will discuss later on (see Eqs~.~(\ref{eq:fplbd}) and (\ref{eq:gsei})), we 
we will find that 
\beq
\sigma _{k}^2=\l(\frac {H ^2}{k^3}\r)\,
\l(1+k^2\eta^2\r);\qquad
{\cal J}_{k}=\l(\frac{k^2} {H ^2\, \eta}\r)\, 
\l(1+k^2\eta^2\r)^{-1}.
\eeq
In such a case, ${\cal J}_{k}(\eta)\to 0, \sigma_k^2\to\infty$  at early times ($\eta\to-\infty$) which corresponds
to a state sharply peaked around the $q$-axis. At late times ($\eta\to0$), however, we have
${\cal J}_{k}(\eta)\to \infty$ with a finite $\sigma_k^2$ which corresponds to a state sharply peaked around
the $p$-axis. In fact, whenever $k\eta\to0$ (corresponding to super Hubble scales), the Wigner function
gets peaked around a classical trajectory. This can be verified more explicitly by studying the classical solution for our problem.
Classically, for the case of exponential inflation, we can write the 
general solution for $q_{\bf k}$ as:
\beq
q_{\bf k}= 2\textrm{Re}\, \left[
- {\cal L}    H\eta\left(
1+\frac{i}{k\eta}\right) \, e^{ik\eta}
\right]
\eeq
where ${\cal L}(k)$ is an arbitrary complex number. Writing 
 ${\cal L}(k)=L(k)\, e^{i\, l(k)}$, this solution becomes
\beq
q_{\bf k}= -2L\,  H \eta\, \cos\l[k\eta+l(k)\r]
+ \frac{2L\,  H }{k}\, \sin\l[k\eta+l(k)\r].
\eeq
The conjugate momentum $p_{\bf k}=\l(a^2\, q_{\bf k}'\r)$ 
corresponding to the above $q_{\bf k}$ is then given by
\beq
p_{\bf k}= -\l(2L\, k/ H \eta\r)\, \sin\l[k\eta+l(k)\r].
\eeq
The trajectory of the system in the phase space is given by
\begin{equation}
\frac{q_{\bf k}}{L} = \frac{H^2\eta}{k^2L} p_{\bf k} \pm 2H\eta \left(1-\frac{H^2\eta^2}{4L^2k^2}p_{\bf k}^2\right)^{1/2}
\label{classtraj}
\end{equation}
At late times (when $\eta\to0$) or at super Hubble scales, we have $(q_{\bf k}/p_{\bf k})\to0$ indicating a trajectory
along the vertical $p$ axis, which is precisely what we get from the Wigner function.  On the other hand, one cannot
naively take the early time limit (when $\eta\to-\infty$) with finite $q_{\bf k},p_{\bf k}$ in the trajectory in Eq.(\ref{classtraj})
since $q_{\bf k}$ becomes imaginary.  One can, however, take the limit of $\eta\to-\infty,p_{\bf k}\to0$ keeping $\eta p_{\bf k}$
constant; in this limit, we obviously get a trajectory along the horizontal $p_{\bf k}=0$ axis, which matches with the analysis based on the Wigner function.  Incidentally, notice that the Hamiltonian for our system has a kinetic energy term
$K\propto p^2/a^2\propto (p\eta)^2$ and a potential energy term $U\propto a^2q^2\propto (q/\eta)^2$. At late times,
potential energy dominates over kinetic energy leading to near classical behaviour peaked around $q=0$; on the other hand, at early times if we let $\eta\to-\infty$ keeping $(p\eta)$ fixed, the kinetic energy remains finite and dominates over the potential energy. This is the quantum regime. (We plan to explore these ideas in detail in a separate publication).

\section{Power law and exponential inflation}\label{sec:plei}

\subsection{Standard initial conditions}

Let us now apply the above formalism to the case of power law inflation.
Power law inflation corresponds to the situation wherein the scale 
factor $a(t)$ grows with $t$ as
\beq 
a(t) = a_{0}\, t^p,
\eeq
where $p > 1$. 
In terms of the conformal time $\eta$, this scale factor can be 
written as
\beq
a(\eta)=\l(-\Hbar \, \eta \r)^{(\gamma+1)},\label{eq:plisf}
\eeq
where $\gamma$ and $\Hbar$ are given by
\beq
\gamma = -\l(\frac{2p-1}{p - 1}\r)
\qquad{\rm and}\qquad
\Hbar = (p - 1) \, a_{0}^{1/p}.
\eeq
Note that $\gamma \le -2$ with $\gamma=-2$ corresponding to exponential 
inflation.
Also, the quantity $\Hbar$ denotes the characteristic energy scale 
associated with inflation and, in the case of exponential inflation, 
it exactly matches the Hubble scale.

On writing $\mu_{k}=(f_{k}/a)$ in the differential 
equation~(\ref{eq:deq}), we find that $f_{k}$ satisfies the following 
equation:
\beq
f_{k}''+\l[k^2-\l(\frac{a''}{a}\r)\r]\, f_{k}=0.\label{eq:deq1}
\eeq
In a Friedmann universe described by the scale factor~(\ref{eq:plisf}), 
the general solution to this differential equation can be written as 
(see, for e.g., Ref.~\cite{as64}, p.~362)
\beq
f_{k}(\eta) 
= \frac{\sqrt{\pi\eta}}{2}\, 
\biggl(e^{-\l(i\pi \gamma/2\r)}\;
H_{-\l(\gamma+\frac{1}{2}\r)}^{(1)}(k\eta)
+ {\cal B}(k)\, e^{\l(i\pi \gamma/2\r)}\; 
H_{-\l(\gamma+\frac{1}{2}\r)}^{(2)}(k\eta)
\biggl),\qquad\label{eq:gsfpl}
\eeq
where $H_{\nu}^{(1)}$ and $H_{\nu}^{(2)}$  are the Hankel 
functions of the first and the second kind, respectively.
The $k$-dependent constant ${\cal B}(k)$ is to be fixed by choosing 
suitable initial conditions for each of the modes. 
For the above solution, it can be easily shown that the Wronskian 
condition~(\ref{eq:wrnskn}) leads to the relation~(\ref{eq:BW}) 
between ${\cal B}(k)$ and $W(k)$.

Let us first briefly review the standard theory (see, for e.g., 
Ref.~\cite{paddy03}) in which the initial conditions are imposed 
on sub-Hubble scales, i.e. when the physical wavelengths 
$(k/a)^{-1}$ of the modes are much smaller than the Hubble radius 
$H^{-1}$.
A natural choice for the initial condition will be the one in which 
each of the oscillators $q_{\bf k}$ is in its ground state at sub-Hubble
scales. 
This condition implies that the wave function (\ref{eq:gswfn}) has 
the following asymptotic form:
\beq
\lim_{\l(k/aH\r)\to \infty} 
\psi_{\bf k}\l(q_{\bf k},\eta\r)
\to \l(\frac{k\, a^2}{\pi}\r)^{1/4}\; 
\exp-\l(\frac{k}{2}\,a^2\, q_{\bf k}^2+i\frac{k}{2}\eta\r)
\eeq
which, in turn, requires that, as $\l(k/aH\r)\to \infty$, we need to 
have $R_{k}\to \l(k\, a^2/2\r)$ and $N_{k}\to \l(k\, a^2/\pi\r)^{1/4}\, 
e^{-ik\eta/2}$.
These conditions can be satisfied provided $ f_{k}\to ({\rm e}^{ik\eta}/
\sqrt{2k})$ as  $\l(k/aH\r)\to \infty$ [cf. Eqs.~(\ref{eq:Rk}), 
(\ref{eq:Nk}) and (\ref{eq:Dk})].
This can be achieved by setting ${\cal B}(k)=0$ in Eq.~(\ref{eq:gsfpl}), 
so that we have
\beq
f_{k}(\eta) = \l(\frac{\sqrt{\pi\eta}}{2}\r)\, 
e^{-\l(i\gamma\pi/2\r)}\, 
H^{(1)}_{-\l(\gamma+\frac{1}{2}\r)}(k\eta)\label{eq:fplbd}
\eeq
and this choice corresponds to what is known as the Bunch-Davies 
vacuum~\cite{bd78}. 
Note that, according to the Eq.~(\ref{eq:BW}), ${\cal B}(k)=0$ 
implies $W(k)=1$.
Therefore, on substituting the above $f_{k}$ in Eq.~(\ref{eq:ps}) 
and imposing the condition of Hubble exit, viz. that $(k/a)=\l(z\, 
H\r)$, we obtain the spectrum of perturbations to be (for a recent 
discussion, see, for e.g., Ref.~\cite{ms03})
\beq
k^3\; {\cal P}_{\Phi}(k)
= C(z) \, \l(\frac{\Hbar^2}{2 \pi ^2}\r) \, 
\l(\frac{k}{\Hbar}\r)^{2(\gamma+2)},\label{eq:psplfr}
\eeq
where $C(z)$ is given by
\beq
C(z) = \l(\frac{\pi}{4}\r)\,
\l\vert H_{-\l(\gamma+\frac{1}{2}\r)}^{(1)}[(\gamma+1)\, z]\r\vert^2\,
\l\vert (\gamma+1)\, z\r\vert^{-(2 \gamma + 1)}.
\eeq
In the limit of $z\to 0$, this expression simplifies to~\cite{ms03}
\beq
C(0)=\l(\frac{2^{-2\l(\gamma+1\r)}}{2\pi}\r)\;
\l\vert\Gamma\l[-\l(\gamma+\frac{1}{2}\r)\r]\r\vert^2.
\eeq
In the case of exponential inflation, corresponding to $\gamma =-2$, 
the perturbation spectrum in Eq.~(\ref{eq:psplfr}) reduces to
\beq
k^3\; {\cal P}_{\Phi}(k)
= \l(\frac{1+z^2}{2}\r)\, \frac{\Hbar^2}{2 \pi^2}\label{eq:psefr}
\eeq
which is a spectrum that is exactly scale invariant. 
(Note that, for exponential inflation, ${\cal H}=H$.)
As we pointed out before, the numerical value of the amplitude
depends on whether we evaluate the expression at $z=1, z=(2\pi)$ or 
as $z\to 0$.

Let us now consider a more general situation with ${\cal B}(k)\neq 0$. 
It is convenient to write
\beq 
{\cal B}(k)=B(k)\, \exp\,\l[i\, b(k)\r],
\label{defB}
\eeq
so that the resulting power spectrum can be expressed as [on assuming
that Hubble exit occurs at $(k/a)=(z\,H)$]
\beq
k^3\; {\cal P}_{\Phi}(k)
= C(z) \, \l(\frac{\Hbar^2}{2 \pi ^2}\r) \, 
\l(\frac{k}{\Hbar}\r)^{2(\gamma+2)}\,
\l[1-B^2(k)\r]^{-1}\;\biggl(1+ B^2(k)
+ 2\,B(k)\, {\rm cos}\l[b(k)+\pi\gamma-2\theta\r]\biggr),
\label{eq:psg}
\eeq
where $\theta$ is the phase of the Hankel function $H_{\nu}^{(1)}$
at Hubble exit, given by the relation
\beq
H_{-\l(\gamma + \frac{1}{2}\r)}^{(1)}[(\gamma+1)z]
=\l\vert H_{-\l(\gamma + \frac{1}{2}\r)}^{(1)}[(\gamma+1)z]\r\vert\; 
e^{i\theta}.\label{eq:theta}
\eeq 
If we write ${\rm cos}\l[b(k)+\pi\gamma-2\theta\r]=d(k)$, then, the 
power spectrum~(\ref{eq:psg}) reduces to
\begin{equation}
k^3\; {\cal P}_{\Phi}(k)
=C(z) \, \l(\frac{\Hbar^2}{2 \pi ^2}\r) \,
\l(\frac{k}{\Hbar}\r)^{2(\gamma+2)}
\l(\frac{1+ B^2(k)+2\,B(k)\,d(k)}{1-B^2(k)}\r),
\label{eq:psgfr}
\end{equation}
where $-1\le d(k)\le 1$.

At super-Hubble scales (i.e. as $z \to 0$), the above power spectrum 
bears a simple relation to the mean number of particles $\langle n_{k}
\rangle$ as given by Eq.~(\ref{eq:nk}).  
We find that they are related as follows:
\beq
k^3\; {\cal P}_{\Phi}(k)
\simeq \l(\frac{\Hbar^2}{\pi ^2}\r)\,
\l(\langle n_{k}\rangle\, \l[\l(\gamma+1\r)\, z\r]^2\r)\,
\l(\frac{k}{\Hbar\, (\gamma+1)\,z}\r)^{2(\gamma+2)}.
\label{eq:psnk}
\eeq
This relation can be easily obtained by using the 
expression~(\ref{eq:calEk}) for ${\cal E}_{k}$ and 
Eq.~(\ref{eq:calEknk}) which relates ${\cal E}_{k}$ to $\langle n_{k} 
\rangle$.
At super-Hubble scales (i.e. as $(k\eta)\to 0$), the general solution 
for $f_{k}$ as given by Eq.~(\ref{eq:gsfpl}) reduces to 
\beq
f_{k}(\eta) \propto
\frac{1}{\sqrt{k}}\, (k\eta)^{(\gamma+1)}.
\eeq
On using this expression, it is straightforward to show that $\mu_{k}'
=(f_{k}/a)' $  is sub-dominant to $\mu_k$ at super-Hubble scales.
Then, from Eqs.~(\ref{eq:calEk}) and (\ref{eq:calEknk}),
we have 
\beq
{\cal E}_{k} \simeq 
\frac{k^2 a^2}{2}\, 
\l(\frac{\vert \mu_{k}\vert^2}{W(k)}\r)
\simeq  \langle n_{k} \rangle\, k.
\eeq
It is then evident from the definition~(\ref{eq:ps}) that the power 
spectrum will be proportional to the mean number of particles at 
super Hubble scales.  
On using the above result, one can easily arrive at the 
relation~(\ref{eq:psnk}), by first imposing the condition for Hubble 
exit [viz. that $\l(k/a\r)=(z\, H)$] and then taking the limit $z\to 0$.
In the limit of exponential inflation (i.e. as $\gamma\to -2$), we find
that $\l[k^3\; {\cal P}_{\Phi}(k)\r] \simeq \l(H^2\, \langle n_{k}\rangle\, 
z^2/\pi^2\r)$ at super-Hubble scales.

In the next section, we shall show that {\it any}\/ modified spectrum 
obtained from a high energy model can be constructed in the standard 
theory by simply choosing suitable forms for the functions~$B(k)$ and
$b(k)$.
But, before we do that, we shall discuss an alternative procedure 
for imposing the initial conditions wherein the initial condition 
for different modes are imposed at different times.

\subsection{Specifying initial conditions when $[k/a(\eta_k)]\simeq 
L_{\rm P}^{-1}$}

In the last section, we chose the quantum state by imposing a 
condition as $(k/aH)\to \infty$ and one may question whether 
we are sure of the short distance, sub-Planck scale physics 
well enough to make this choice. 
An alternative procedure, discussed extensively in the 
literature~\cite{minimal}, attempts to address this question by 
choosing to impose the initial conditions {\it for the different 
oscillators at different times}.\/  
Specifically, one chooses the initial condition for the oscillator 
$q_{\bf k}$ at a time $\eta_{k}$ such that  $[k/a(\eta_k)]\simeq 
L_{\rm P}^{-1}$, where $L_{\rm P}$ is the Planck length. 
The initial state of the oscillator uniquely determines the function 
${\cal B}(k)$ and the corresponding power spectrum can then be 
constructed  using Eq.~(\ref{eq:psg}). 
However, it should be emphasized here that, apart from the fact 
that it has to be consistent with the observations, there exists 
no restrictions on the initial state to be chosen at $\eta_{k}$.

To illustrate the point, let us consider the  case of 
exponential expansion for which $a(\eta)=(-{\cal H}\eta)^{-1}$. 
Let us assume that the oscillator $q_{\bf k}$ is in the ground state
at a given time, say, $\eta_{i}$, which then requires that, at 
this instant, $R_{k}=-\l(i\, a^2/2\r)\l(\mu_{k}'/ \mu_{k}\r)
=\l(k\, a^2/2\r)$.
For the case of exponential expansion, the general solution for 
$f_{k}$ as given by Eq.~(\ref{eq:gsfpl}) can be expressed in 
terms of simple functions as follows (see, for instance, 
Ref.~\cite{as64}, pp.~437--438):
\beq
f_{k}(\eta)
= \frac{1}{\sqrt{2k}}\, 
\biggl[\l(1+\frac{i}{k\eta}\r)\, e^{ik\eta}
+\; {\cal B}(k)\, \l(1-\frac{i}{k\eta}\r)\, e^{-ik\eta}\biggr].
\label{eq:gsei}
\eeq
On imposing the condition $R_{k}=\l(k\, a^2/2\r)$ at $\eta_{i}$, 
we can determine the function ${\cal B}(k)$ to be
\beq
{\cal B}(k)=\l(1+2i\, k\eta_{i}\r)^{-1}\, e^{2ik\eta_{i}}.
\label{eq:BkD}
\eeq
Our original choice of the Bunch-Davies vacuum corresponds to 
assuming $\eta_i\to-\infty$ so that ${\cal B}(k)$ vanishes for 
all~$k$. 
The modified procedure will be to choose $\eta_{i}$ differently for 
different $k$ by imposing the initial condition at, say, $[k/a(\eta_k)]
=L_{\rm P}^{-1}$. 
For exponential inflation, this translates to $(k\, \eta_{k})= 
-({\cal H}\, L_{\rm P})^{-1} \equiv -\xi^{-1}$, a constant. 
For $\eta_{i}=\eta_{k}$, the ${\cal B}(k)$ above reduces 
to~\cite{minimal,mb03}
\beq
{\cal B}(k)=\l[1-(2i/\xi)\r]^{-1}\, e^{-(2i/\xi)}\label{eq:Bkdpvi}
\eeq
which is a {\it constant independent of $k$}.\/ 
Thus the $k$-dependence of the power spectrum remains unchanged when 
the initial condition on the $k$-th oscillator is imposed at a time 
such that $[k/a(\eta_{k})]={\rm constant}$. 
The amplitude of the power spectrum, of course, gets scaled by a 
$k-$independent factor; this is of no observable consequence since 
we do not know how to obtain the amplitude from a first principle 
theory anyway.

The power spectrum corresponding to the above ${\cal B}(k)$ when 
evaluated at Hubble exit, say, when $(k/a)=\l(z\, H\r)$, is given by
\beq
k^3\; {\cal P}_{\Phi}(k)
= \l(\frac{1+z^2}{2}\r) \l(\frac{\Hbar^2}{2\pi^2}\r)
\l[1+\frac{\xi^2}{2}
+\l(\frac{(z^2-1)\, \xi^2-4z\, \xi}{2\, (1+z^2)}\r) 
{\rm cos}\l(\frac{2}{\xi} -2z\r) 
+ \l(\frac{(z^2-1)\, \xi + z\,\xi^2}{1+z^2}\r)
{\rm sin}\l(\frac{2}{\xi}-2z\r)\r].\label{eq:Dpsg}
\eeq
This expression is exact and shows that the spectrum is strictly 
scale invariant, but is modified from the original result [viz.
Eq.~(\ref{eq:psefr})] by a multiplicative factor independent of $k$. 
This factor depends on two parameters: (i)~$\xi=\l({\cal H}\, 
L_{\rm P}\r)$ which measures the energy scale of inflation relative 
to the Planck scale and (ii)~$z$ which is the ratio of the Hubble 
radius and the physical wavelength of the perturbation. 
As we explained before, we can evaluate the power spectrum either at 
$z=1$ or at $z=(2 \pi)$. 
For $z=1$, the spectrum~(\ref{eq:Dpsg}) reduces to
\beq
k^3\; {\cal P}_{\Phi}(k)
=\l(\frac{\Hbar^2}{2\pi^2}\r)\,\l[1+\frac{\xi^2}{2}
-\xi\, {\rm cos}\l(\frac{2}{\xi} -2\r) 
+ \frac{\xi^2}{2}\, {\rm sin}\l(\frac{2}{\xi}-2\r)\r],
\eeq
while, for $z=(2\pi)$, we get
\beq
k^3\; {\cal P}_{\Phi}(k)
= \l(\frac{1+4\pi^2}{2}\r)\, 
\l(\frac{\Hbar^2}{2\pi^2}\r)\,\l[1+\frac{\xi^2}{2}
+\l(\frac{(4\pi^2-1)\, \xi^2-8\pi\, \xi}{2\, (1+4\pi^2)}\r)\, 
{\rm cos}\l(\frac{2}{\xi}\r) 
+ \l(\frac{(4\pi^2-1)\, \xi + 2\pi\xi^2}{1+4\pi^2}\r)\, 
{\rm sin}\l(\frac{2}{\xi}\r)\r].
\eeq 
The numerical value of these modified amplitudes will depend on 
the parameter $\xi$ which is expected to be very small compared 
to unity for GUT scale inflation.
For $\xi\ll 1$, one can easily obtain the leading order terms of 
polynomial expressions involving $\xi$, but determining $\cos(1/\xi)$ 
and $\sin(1/\xi)$ for $\xi\ll 1$ requires care. 
Since, one can easily replace $(1/\xi)$ by, say, $[(1/\xi)+(\pi/2)]$, 
in the arguments of trigonometric functions to the leading order, these 
expressions are intrinsically ambiguous. 
For $\xi\ll 1$ and $z=1$, we get
\beq
k^3\; {\cal P}_{\Phi}(k)
\simeq \l(\frac{\Hbar^2}{2\pi^2}\r)\,
\l[1- \xi\, F\l(2/\xi\r)\r],
\eeq
where $F$ is a rapidly oscillating cosine function. 
(Note that, if one evaluates the spectrum at super Hubble scales, 
i.e. as $z\to 0$, then, instead of the cosine, one gets a sine 
function~\cite{minimal}.)
Similarly, when $\xi\ll 1$, for $z=(2\pi)$, we have
\beq
k^3\; {\cal P}_{\Phi}(k)
\simeq 
\l(\frac{1+4\pi^2}{2}\r)\,
\l(\frac{\Hbar^2}{2\pi^2}\r)\,\l[1
-\l(\frac{4\pi\,\xi}{1+4\pi^2}\r)\, F(2/\xi) 
+ \l(\frac{(4\pi^2-1)\, \xi}{1+4\pi^2}\r)\, G(2/\xi)\r],
\eeq
where $G$ is a sine function and $F$, as above, is a cosine function.
In each of these cases, a different choice for the sub-leading phase 
in the argument of the trigonometric function can make cosine into 
sine and vice versa.
Hence, only the profile of the oscillating functions are of relevance 
in the limit of $\xi\ll 1$, though the full expressions are often 
quoted in literature.

Note that the quantum state with the choice of ${\cal B}(k)$ in 
Eq.~(\ref{eq:Bkdpvi}) is a state with the initial condition for 
all the oscillators specified at a given moment of time.
Thus the above analysis maps the prescription of specifying the 
quantum state for different oscillators at different times to 
specifying the initial condition at a given time. 
We can now explore the physical content of this quantum state at 
any given $\eta$ in terms of, for e.g., the mean occupation number 
$\langle n_{k}\rangle$ in the instantaneous harmonic oscillator
states.
On using the expressions~(\ref{eq:Gamma}), (\ref{eq:nk}),
(\ref{eq:gsei}) and (\ref{eq:Bkdpvi}), we find the particle content 
of this quantum state to be
\beq
\langle n_{k}\rangle = \l(\frac{1}{4k^2\eta^2}\r)\, 
\biggl[1+ \frac{\xi^2}{2}+ \xi^2\, k^{2}\eta^2
-\frac{\xi}{2}\l(\xi-4\, k\eta\r)\, 
{\rm cos}\l(\frac{2}{\xi}+2k\eta\r)
- \xi\, \l(1+\xi\, k\eta\r)\, 
{\rm sin}\l(\frac{2}{\xi}+2k\eta\r)\biggr].\label{eq:nkD}
\eeq
This expression has several interesting features. 

First, let us consider very early times (i.e. as $\eta\to -\infty$) 
or very short wavelengths (i.e. as $k\to\infty$). 
It is precisely this limit which was considered uncertain due 
to trans-Planckian effects which motivated imposing the initial 
condition for different modes at different times; therefore, it 
is this limit which is of some interest to explore, to understand 
what kind of {\it effective}\/ quantum state at $\eta=$ constant 
$\to -\infty$ will lead to the vacuum state for the $k$-th mode 
when $(k\eta)=-\xi^{-1}$. 
We see from Eq.~(\ref{eq:nkD}) that the mean occupation number has 
two sets of contributions. 
The first three terms in Eq.~(\ref{eq:nkD}) is secular, while the last 
two terms are oscillatory. 
The secular term increases monotonically from $\langle n_{k}\rangle
=\l(\xi^2/4\r)$ at $\eta=-\infty$. 
Thus, in the trans-Planckian limit (i.e. as $\eta\to -\infty$ or  
as $k\to\infty$) all the modes have the same mean excitation 
$(\xi^2/4)$. 
(The oscillatory terms do not contribute in this limit.)  

Second, we do know that $\langle n_{k}\rangle$ must vanish for 
$(k\eta)=-\xi^{-1}$, since this is the condition we used to choose 
this state. 
As can be directly verified, this is indeed true for the expression 
in Eq.~(\ref{eq:nkD}), but occurs because of a cancellation between 
the secular and the oscillatory terms. 
At later times, the secular terms dominate the oscillatory terms and 
$\langle n_{k}\rangle$ increases on the average with superimposed 
oscillations. 
(We stress the fact that $\langle n_{k}\rangle$ is computed in terms 
of the instantaneous harmonic oscillator modes; one should not think 
of them as particles which are produced with respect to the in-out
states.) 
Thus, for each of the modes, we start with $\langle n_{k}\rangle
=(\xi^2/4)$ as $\eta\to-\infty$, evolve to $\langle n_{k}\rangle=0$ 
at $(k\eta)=-\xi^{-1}$ and grow to 
\beq
\langle n_{k}\rangle = \l(\frac{1}{4z^2}\r)\, 
\biggl[1+ \frac{\xi^2}{2}+ \xi^2\, z^2
-\frac{\xi}{2}\l(\xi+4\,z\r)\, 
{\rm cos}\l(\frac{2}{\xi}-2z\r)
- \xi\, \l(1-\xi\, z\r)\, 
{\rm sin}\l(\frac{2}{\xi}-2z\r)\biggr]
\eeq
at Hubble exit. 
We see that the alternative prescription of specifying the initial 
condition when the physical wavelength of the mode is comparable to 
the Planck length (which sounds reasonable at first sight) is 
equivalent to assuming a quantum state at a $\eta$= constant 
hypersurface, with very specific properties. 
It can be realized only if unknown physical effects of the 
trans-Planckian sector acts in a particular manner to populate the 
modes  with a specific prescription. 
It is far from clear whether this will be possible in a generic context.

\section{Mimicking the trans-Planckian effects}\label{sec:mtpe}

We shall now turn to the question of choosing an initial state such 
that a given power spectrum of perturbations is reproduced.
Let us  assume that, in the case of power-law inflation, a 
`trans-Planckian' theory leads to the following spectrum:
\beq
\left[k^3\; {\cal P}_{\Phi}(k)\right]_{\rm M}
= C(z)\,\l(\frac{\Hbar^2}{2\pi^2}\r)\,
\l(\frac{k}{\Hbar}\r)^{2 (\gamma + 2)}\, M(k),\label{eq:psplm}
\eeq
where $M(k)$ is the modification factor with, of course, $M(k)\ge 0$ 
for all~$k$.
(Alternatively, let us suppose that a future CMB observation leads 
to such a spectrum, starting a hectic flurry of theoretical activity 
to explain it!)
Evidently, the modified spectrum~(\ref{eq:psplm}) can be constructed 
from the spectrum~(\ref{eq:psgfr}) in the standard theory, provided 
we can find a positive definite function~$B(k)$ that satisfies the 
condition
\beq
\l[1+ B^2(k)+2\, B(k)\, d(k)\r]=\l[1-B^2(k)\r]\, M(k).
\label{eq:cndtn}
\eeq
This is a quadratic equation in $B(k)$ for each value of $k$ and, 
if we choose $d(k)=\pm 1$, then, we find that the non trivial and 
positive definite roots for $B(k)$ can be expressed in terms of 
$M(k)$ as follows:
\beq
B(k)=\frac{\vert M(k)-1\vert}{M(k)+1}.\label{eq:BkMk}
\eeq
Given a $M(k)$, this expression then allows us to reproduce the 
modified spectrum in the standard theory. 
However, note that the $B(k)$ we have constructed above is not unique 
as it corresponds to a particular value of the phase $b(k)$ in Eq.~(\ref{defB})
such that $d(k)=\pm 1$.
Clearly, a whole {\it class}\/ of such functions can be constructed, with 
different choices for $d(k)$, but, as we shall see, this particular choice 
is sufficient for our purpose.

As we had mentioned in Section~\ref{sec:intro}, the Planck scale 
modifications to the standard inflationary perturbation spectrum 
have been obtained in the literature by considering various high 
energy models for the scalar field~\cite{tplp}--\cite{mb03}.
Most of these modified spectra deviate from the standard scale 
invariant spectrum at the ultra-violet end.
On the other hand, the lower power in the quadrupole and the octopole 
moments of the CMB as measured by WMAP~\cite{WMAP}---if it survives 
further releases of WMAP data!---requires a suppression of power in 
the infra-red end of the spectrum~\cite{blwe03,lqf}. 
Such spectra have also been obtained in certain high energy models 
of the scalar field~\cite{nccmb,ss04}.  
It is  interesting to determine how these different modified spectra 
can be constructed by choosing suitable initial conditions on the 
scalar field in the standard theory.

In the following subsections, we shall explicitly construct the 
function $B(k)$ for four modified power spectra that have either 
been proposed to fit the observational data or have been 
obtained from a high energy model of the scalar field.

\subsection{Modified spectrum I}

Recently, the following spectrum was obtained in a power law
inflationary scenario using a Lorentz invariant modified
theory~\cite{ss04}:
\beq
\left[k^3\; {\cal P}_{\Phi}(k)\right]_{\rm M}
= C(1)\,\l(\frac{\Hbar^2}{2\pi^2}\r)\,
\l(\frac{k}{\Hbar}\r)^{2 (\gamma + 2)}\; 
\l[1 - {\bar C}\,\l(\frac{\Hbar}{k_{c}}\r)\,
\l(\frac{k}{\Hbar}\r)^{(\gamma + 2)}\r],\label{eq:psmli}
\eeq
where $k_{\rm c}$ denotes the high energy scale, ${\bar C}$ is 
given by 
\beq
{\bar C} = \l[2\, C(1)\, \l(\gamma +1\r)^{3(\gamma + 1)}\r]^{-1}
\eeq
and it is assumed that $10^{-5} \lesssim \l(\Hbar/k_{\rm c}\r) 
\lesssim 10^{-3}$.
(However, it should be mentioned here that this spectrum is not 
valid for arbitrarily small values of $k$~\cite{ss04}.)
In fact, a similar spectrum have also been obtained in non-commutative 
models of inflation~\cite{nc,nccmb}.
These spectra exhibit a suppression of power at the large scales, a 
feature that could be relevant to the low quadrupole moment observed 
in the CMB~\cite{blwe03}.
It is straightforward to see that, for $z=1$, the function $B(k)$ 
corresponding to the modified spectrum~(\ref{eq:psmli}) is given by
\beq
B(k)=\l[\l(\frac{2}{{\bar C}}\r)\, \l(\frac{\Hbar}{k_{c}}\r)^{-1}\,
\l(\frac{k}{\Hbar}\r)^{-(\gamma + 2)}-1\r]^{-1}.
\eeq
For $\gamma<-2$, we have  $B\to k^{(\gamma+2)}\to 0$ as $k\to \infty$.
This implies that, while, towards the infra-red end, the initial 
state is different from the Bunch-Davies vacuum, the Bunch-Davies 
vacuum structure is retained at the ultraviolet end. 

\subsection{Modified spectrum II}

Another modified primordial spectrum that has been {\it proposed}\/ 
(see, for e.g., Refs.~\cite{lqf}) in order to account for the low 
quadrupole moment observed in the CMB, is the following:
\beq
\left[k^3\; {\cal P}_{\Phi}(k)\right]_{\rm M}
=A\, k^{\l(n_{\rm s}-1\r)}\, \l[1-e^{-(k/k_{1})^\alpha}\r],
\label{eq:cmbqps}
\eeq
where $A$ and $n_{\rm s}$ are the scalar amplitude and index of the 
standard spectrum.
The pivot scale $k_{1}$ and the constant $\alpha$ (which turns out to 
be a positive number of the order of unity) are chosen to fit the CMB 
data.
If we now assume that amplitude $A$ and the index $n_{\rm s}$ 
of the above modified spectrum are the same as those  in the case of 
power-law inflation in the standard theory [cf.~Eq.~(\ref{eq:psplfr})], 
then, we find that $B(k)$ is given by
\beq 
B(k)=\l[2\, e^{(k/k_{1})^\alpha}-1\r]^{-1}.
\eeq
Note that $B(k)\to 1$ as $k\to 0$ and $B_{k}\simeq e^{-(k/k_{1})^\alpha}
\to 0$ for large $k$.
Once again, the vacuum structure at the ultraviolet end is not 
modified.

\subsection{Modified spectrum III}\label{subsec:mtpps}

Another modified spectrum whose effects on the CMB has also been 
analyzed is the following spectrum \cite{minimalcmb}
\beq
\left[k^3\; {\cal P}_{\Phi}(k)\right]_{\rm M}
=\l(\frac{\Hbar^2}{2\, \pi^2}\r)\!
\biggl(1-\xi\, \l(\frac{k}{k_{2}}\r)^{-\epsilon}\, 
{\rm sin}\biggl[\frac{2}{\xi}\! \l(\frac{k}{k_{2}}\r)^{\epsilon}
\biggr]\biggr),
\label{eq:mtpps}
\eeq
where $\xi\simeq 10^{-3}$, $\epsilon\simeq 10^{-2}$ and $k_{2}$ 
is a pivot scale.
In order to match the leading term in the spectrum, let us assume that 
the inflating background undergoes exponential expansion and that the
spectrum is evaluated at $z=1$.
It is then straightforward to construct $B(k)$ for the above 
spectrum and it is given by
\beq
B(k)=\l(\frac{\xi\, \l(k/k_{2}\r)^{-\epsilon}\, 
\l\vert{\rm sin}\l[\l(2/\xi\r)\l(k/k_{2}\r)^{\epsilon}
\r]\r\vert}{2-\xi\, \l(k/k_{2}\r)^{-\epsilon}\, 
{\rm sin}\l[\l(2/\xi\r) \l(k/k_{2}\r)^{\epsilon}\r]}\r).
\eeq

\subsection{Modified spectrum IV}

A more general modification of spectrum one can envisage is a 
spectrum which has corrections at both the infra-red and the 
ultra-violet ends.
If we now assume that the standard spectrum is modified in  the 
infra-red as in Eq.~(\ref{eq:cmbqps}) and  has the same 
correction at the ultra-violet end as in Eq.~(\ref{eq:mtpps}), 
then the complete spectrum will be given by
\begin{widetext}
\beq
\left[k^3\; {\cal P}_{\Phi}(k)\right]_{\rm M}
=A\, k^{\l(n_{\rm s}-1\r)}\, \biggl(1-e^{-(k/k_{1})^\alpha}
-\xi\, \l(\frac{k}{k_{2}}\r)^{-\epsilon}\!
{\rm sin}\l[\frac{2}{\xi} \l(\frac{k}{k_{2}}\r)^{\epsilon}\r]\biggr),
\label{eq:combops}
\eeq
\end{widetext}
where $A$ and $n_{\rm s}$ are the scalar amplitude and index of the 
standard spectrum and  $k_{1}$ and $k_{2}$ are pivot scales such that 
$k_{1}\ll k_{2}$.
Also, as in the earlier case, let $\alpha$ be a positive constant of 
the order of unity. 
The corresponding $B(k)$ can be easily obtained to be
\beq 
B(k)\!= \! \Biggl(\frac{\l\vert e^{-(k/k_{1})^\alpha}
+\xi\, \l(k/k_{2}\r)^{-\epsilon} 
{\rm sin}\l[(2/\xi)\! \l(k/k_{2}\r)^{\epsilon}\r]\r\vert} 
{2-e^{-(k/k_{1})^\alpha}-\xi\, \l(k/k_{2}\r)^{-\epsilon}
{\rm sin}\l[(2/\xi)\! \l(k/k_{2}\r)^{\epsilon}\r]}\Biggr).
\eeq

These examples demonstrate the fact that the modification of the spectrum 
due to `trans-Planckian considerations' is degenerate with the choice of 
the initial quantum state. 
Without further input, such as an assumption for the choice of initial 
quantum state, observations cannot distinguish between these two physical 
effects.

\section{Modified spectra and back reaction}\label{sec:msbr}

An issue that remains unresolved in obtaining the modified spectra 
is whether the conditions that lead to modifications of the standard 
spectrum will also lead to a large back reaction on the inflating 
background.
In particular, will the energy in the quantum field dominate the
inflaton energy thereby, possibly, terminating inflation?
The approach we have adopted here allows us to address this issue
along the following manner.

Since the modified spectra from a high energy theory can be 
obtained from the standard theory with a suitable choice of 
initial conditions on the modes, we are probably justified in using 
these modes to evaluate the energy density of the quantum 
scalar field (for an earlier discussion on this point, see 
the first reference listed under Ref.~\cite{br}).
In what follows, we shall show that the energy density of the 
quantum scalar field {\it above}\/ the Bunch-Davies vacuum is 
finite only if $\vert M(k)-1\vert$ decays faster than $k^{-2}$ 
for large $k$ and $M(k)$ goes as $k^{\lambda}$ with $-2<\lambda 
<2$ for small $k$.
Also, we shall restrict our attention to the easily tractable 
case of exponential inflation.

Recall that the energy density of the quantum scalar field is 
given by Eq.~(\ref{eq:ed}) with the quantity ${\cal E}_{k}$ to 
be evaluated using Eq.~({\ref{eq:calEk}).
In the case of exponential inflation, as we had mentioned earlier, 
the general solution for $f_{k}$ can be expressed in terms of simple 
functions [cf.~Eq.~(\ref{eq:gsei})].
On substituting the solution (\ref{eq:gsei}) in the 
expression~(\ref{eq:calEk}), we find that the energy density per 
mode of the quantum field is given by
\beq
{\cal E}_{k} = \l(\frac{1}{4k\eta^2}\r)\, 
\l[1-B^2(k)\r]^{-1}\biggl(\l[1+B^2(k)\r]\, \l(1+2\, k^{2}\eta^2\r)
-2 B(k)\, {\rm cos}[b(k)-2k\eta]
+ 4B(k)\, (k\eta)\, {\rm sin}[b(k)-2k\eta]\biggr).
\eeq
In the Bunch-Davies vacuum, which corresponds to ${\cal B}(k)=0$, 
this expression for ${\cal E}_{k}$ reduces to 
\beq
{\cal E}_{k}^{\rm BD}
=\frac{k}{2}+\frac{1}{2k\eta^2}\label{eq:calEkBD}
\eeq
and it known that the corresponding energy density (when 
regularized by point-splitting) is given by~\cite{bd78,bd82}
\beq
\rho_{\rm BD} = -\l(\frac{29}{960\, \pi^2}\r)\, {\cal H}^4.
\eeq 
This energy density is much smaller than the energy density in the 
classical field that drives inflation.
The energy density of the scalar field {\it above}\/ the Bunch-Davies 
vacuum is then given by Eq.~(\ref{eq:ed}) with ${\cal E}_{k}$ replaced 
by ${\bar {\cal E}}_{k}$, where
\beq
{\bar {\cal E}}_{k} = {\cal E}_{k}-{\cal E}_{k}^{\rm BD}\nn\\
= \l(\frac{1}{2k\eta^2}\r)\, 
\l[1-B^2(k)\r]^{-1}
\biggl(B^2(k)\,  \l(1+2\, k^{2}\eta^2\r)
- B(k)\, {\rm cos}[b(k)-2k\eta]
+\; 2 B(k)\, (k\eta)\, {\rm sin}[b(k)-2k\eta]\biggr).
\label{eq:calEbark}
\eeq

In order to understand the behavior of the above ${\bar {\cal E}}_{k}$ 
for large $k$, we can ignore the terms containing the sine and the 
cosine functions as they will oscillate rapidly in this limit. 
On neglecting these terms and, on making use of the 
relation~(\ref{eq:BkMk}) between $B(k)$ and $M(k)$, we find that, for
large k, the above expression for ${\bar {\cal E}}_{k}$ can be written 
in terms of $M(k)$ as follows:
\beq
{\bar {\cal E}}_{k} 
\simeq \l(\frac{\l[M(k)-1\r]^{2}}{4\, M(k)}\r)\, k.
\label{meq}
\eeq
For the energy density $\rho$ corresponding to this ${\bar {\cal E}}_{k}$ 
to be finite in the ultraviolet limit, the integral of $\l(k^2\, {\bar 
{\cal E}}_{k}\r)$ over $k$ should converge at large~$k$. 
It is easy to see that this condition cannot be satisfied if $M(k) \propto 
k^{\nu}$ for any value of $\nu$. 
For all values of $\nu$, this expression varies as $k^{(3+\nu)}$ at 
large $k$ and hence the integral is divergent. 
To obtain a finite result, we need $M(k)\to 1$ at large $k$.
If we now assume that $M(k)\propto \l(1\pm k^{-\delta}\r)$ for large 
$k$, it is then clear that the energy density corresponding to the 
above ${\bar {\cal E}}_{k}$ will converge only if $\delta>2$. 
Thus, the deviations from the standard spectra should die down faster 
than $k^{-2}$ for large $k$.

One can also investigate the infrared limit in a similar fashion. 
As $k\to 0$, we find that the leading divergence arises due 
to the first term in the expression (\ref{eq:calEbark}) for 
${\bar {\cal E}}_{k}$ so that, in this limit, we have
\beq
{\bar {\cal E}}_{k} 
\propto \l(\frac{\l[M(k)-1\r]^{2}}{M(k)\, k}\r).
\eeq 
If we now assume that $M(k)\propto k^\lambda$ as $k\to 0$, then the 
finiteness of the energy density $\rho$ corresponding to this
${\bar {\cal E}}_{k}$ requires that $-2< \lambda <2$. 
It is therefore possible to enhance or reduce power in the infrared 
limit within a range and still maintain finite energy density. 
Modifications of the form $\l[M(k)-1\r]\propto k^\varepsilon$ with 
$\varepsilon >0$ are also allowed and there are no restrictions 
on $\varepsilon$ in this case. 
(We stress that the finiteness of $\rho$ is a necessary condition 
for ignoring back reaction, but it is not sufficient. 
The latter will require comparing the energy density in the quantum
field with the background energy density which is difficult to do 
without assuming a specific model.)
Clearly, amongst the four modified spectra that we have considered in 
the last section, only the second spectrum (provided $\alpha < 2$)
will lead to a finite energy density for the case of exponential 
inflation.

Our expression in Eq.~(\ref{meq}) shows that a constant $M(k)$ 
independent of $k$ leads to a divergent energy density above the 
Bunch-Davies vacuum. 
In particular, the state obtained by giving initial conditions for 
each of the modes at $[k/a(\eta_{k})]$=constant (leading to a 
${\cal B}(k)$ that is independent of $k$), produces a divergent 
contribution to the energy density and hence is suspect as a valid 
quantum state. 
It is sometimes argued in the literature that this state has the
same energy density as the Bunch-Davies vacuum along the following 
lines: When ${\cal B}$ is independent of $k$, we are dealing with 
mode functions of the form $(f_{k}+{\cal B}\, f^*_{k})$, with 
$f_{k}$ given by Eq.~(\ref{eq:gsei}), which belong the set of so 
called $\alpha$-vacua~\cite{alphav}. 
It is possible to construct a regularization scheme such that the 
divergences which arise at the coincidence limit of the Greens 
function in this case, is the same as that in the case of ${\cal B}
=0$. 
Such a subtraction will lead to $\rho=\rho_{BD}$ for these states. 
In our approach, this is equivalent to ignoring the contributions 
in Eq.~(\ref{meq}) when $M(k)=M_0$ is a constant different from unity. 
Since this leads to ${\bar {\cal E}}_{k}\propto k$,
one can think of $M_0$ as the (constant) occupation number 
$\langle n_{k}\rangle$ for all $k$; the regularization involves 
ignoring all these `particles' in measuring the energy density.

We believe this argument is spurious for two reasons. 
Firstly, a transparent and direct discussion in terms of harmonic 
oscillators presented above gives a different result showing that 
$M(k)=M_0$= constant leads to a divergent energy density. 
So, at the least, the results depend on the explicit regularization
procedure adopted.
Secondly, in a realistic model, we will have to deal with a weak 
dependence of ${\cal B}$ on $k$ rather than no dependence. 
Then, the argument based on $\alpha$-vacua will no longer hold, 
irrespective of how weak this dependence is. 
The analysis given above, however, is completely transparent and 
clear.

\section{Discussion}\label{sec:dscssn}

The fact that the measurements of the CMB anisotropies strongly 
indicate a primordial spectrum that is nearly scale invariant 
already implies that the initial state of the quantum scalar field 
is a state that is `very close' to the Bunch-Davies vacuum.
Clearly, sufficiently precise measurements of the anisotropies
in the CMB can provide us with the form the inflationary perturbation 
spectrum to a good accuracy.
However, this information can at most help us determine the initial 
state of the quantum scalar field in the standard theory and it is 
not sufficient to aid us in discriminating between the various Planck 
scale models of matter fields.
The reason being that, given a perturbation spectrum, we should be
able obtain the spectrum from {\it any}\/ high energy model of the
matter field by simply choosing a suitable state in the modified
theory just as we had done in the standard theory.

The above result can be obtained either in the Heisenberg picture or 
in the Schrodinger picture since these descriptions are mathematically
equivalent. 
We have, however, adapted the Schrodinger picture since the problem 
we are discussing is essentially that of a harmonic oscillator with 
a time dependent frequency for which the intuition available in
the Schrodinger picture is of some value. 
Our discussion clearly shows that virtually any power spectrum which 
is either observed in future or suggested by phenomenological models, 
can be reproduced by a suitable choice of the quantum state. 
Evidently, without additional assumptions one cannot disentangle the 
dynamics from the initial conditions and, hence, the CMB observations 
alone cannot act as a discriminator between different theoretical models. 

There are several possible avenues for future work arising from this 
discussion.
One particularly interesting question will be the evolution of the 
quantum state of the universe into the future. 
Several recent observations \cite{snmap} (as well as not so recent 
observations, see Refs.~\cite{early}) suggest that the universe has 
just entered an accelerating phase dominated by dark energy with an 
equation of state $P\approx -\rho$. 
While the nature of this dark energy is unclear, it seems likely that 
at least at sufficiently large scales it will act like a cosmological 
constant leading to a late time de Sitter phase~\cite{tptrc}. 
It will be interesting to study the late time evolution of the quantum 
wave function of the scalar field. 
The de Sitter phase in the future should lead to its own thermal 
fluctuations with a characteristic temperature and it will be 
interesting to see how that can emerge.

Another issue of interest is the study of correlations across the 
horizon in the case of de Sitter spacetime. 
It has been shown that the quantum entanglement of modes across the 
horizon can lead to a holographic description of gravity~\cite{tpseries}. 
The effects of this entanglement are easy to calculate when de Sitter 
spacetime is described in the static coordinates. 
On the other hand, the time dependent Gaussian state used in this 
paper is more naturally tuned to the Friedmann coordinates of the 
de Sitter spacetime. 
It will be worthwhile to relate these two descriptions and understand 
how the correlations across the horizon arises in the time dependent 
background. 

Finally, the description in terms of the wave function can be easily 
translated to one in terms of the path integral kernel using the 
Feynman-Kac formula.
This will allow one to provide a purely path integral derivation of 
the results presented in this paper. 
All these issues are currently under investigation. 

\appendix
\section{Evaluating the power spectrum}\label{app:eps}

Recall that the quantum state $\psi_{\bf k}$ of the time dependent 
oscillator $q_{\bf k}$ was described by the Gaussian wave 
function~(\ref{eq:gswfn}).
On substituting the wave function~(\ref{eq:gswfn}) in the Schrodinger 
equation~(\ref{eq:seq}) and equating the coefficients of $q_{\bf k}$ and
$q_{\bf k}^2$, we find that $N_{k}$ and $R_{k}$ satisfy the following
differential equations:
\br
i\, N_{k}'&=& \frac{R_{k}\, N_{k}}{a^2},\label{eq:deNk}\\
i\, R_{k}'
&=& \frac{2R_{k}^{2}}{a^2}
-\frac{k^{2} a^2}{2},\label{eq:deRk}
\er
where the prime, as before, denotes differentiation with respect 
to $\eta$.
Let us now introduce a new quantity $\mu_{k}(\eta)$ through the 
relation~\cite{paddy03,tpgswfn,ss}
\beq
R_{k}=-\l(\frac{i\, a^2}{2}\r)\, \l(\frac{\mu_{k}'}{\mu_{k}}\r).
\label{eq:Rk}
\eeq
On substituting this expression in the above differential equation 
for $R_{k}$, we find that $\mu_{k}$ satisfies the differential 
equation~(\ref{eq:deq}) which is the same as the classical equation 
of motion satisfied by the classical oscillator variable $q_{\bf k}$.
Also, in terms of $\mu_{k}$,  the differential equation~(\ref{eq:deNk})
can be integrated to obtain
\beq
N_{k}= \l(\frac{D(k)}{\sqrt{\mu_{k}}}\r),\label{eq:Nk}
\eeq
where $D(k)$ is a $k$-dependent constant determined by the 
normalization condition~(\ref{eq:NkRk}). 
Note that the differential equation satisfied by $\mu_{k}$ [viz. 
Eq.~(\ref{eq:deq})] implies the Wronskian condition 
\beq
\l(\mu_{k}\, {\mu_{k}'}^{\ast} - \mu_{k}'\, \mu_{k}^{\ast}\r)
=-\l[i\, W(k)/a^2\r],\label{eq:wrnskn}
\eeq
where $W(k)$ is a $k$-dependent constant.
The relations~(\ref{eq:Rk}) and (\ref{eq:Nk}) along with the above 
Wronskian condition and the normalization condition~(\ref{eq:NkRk})
determine $D(k)$ to be
\beq
D(k)=\l(\frac{W(k)}{2\pi}\r)^{1/4}.\label{eq:Dk}
\eeq 
Thus, the solution to classical equation of motion allows us 
to construct the wave function for the corresponding quantum 
problem~\cite{dr94,paddy03,tpgswfn,ss}.
However, it should be pointed out here that, while $q_{\bf k}$ is real, 
$\mu_{k}$, in general, is a complex quantity.

The power spectrum of the perturbations~(\ref{eq:psdfntn}) is given by
\beq
k^3\; {\cal P}_{\Phi}(k)
= \frac{k^3}{2\pi^{2}}
\int\limits_{-\infty}^{\infty} dq_{\bf k}\, 
\vert \psi_{\bf k}\vert^2\, q_{\bf k}^2
=\frac{k^3\, \vert N_{k}\vert^2}{2\pi^{2}}\,
\int\limits_{-\infty}^{\infty} dq_{\bf k}\;
q_{\bf k}^{2}\; e^{-\l[\l(R_{k}+R_{k}^{\ast}\r)q_{\bf k}^2\r]},
\label{eq:psplintgrl}
\eeq
where we have substituted the expression~(\ref{eq:gswfn}) for the 
wave function $\psi_{\bf k}$.
On carrying out the integral over $q_{\bf k}$ in the above equation and 
making use of the relation~(\ref{eq:NkRk}) and the Wronskian 
condition~(\ref{eq:wrnskn}), we obtain the power spectrum to be
\beq
k^3\; {\cal P}_{\Phi}(k)
=\frac{k^3}{2\pi^{2}}\, 
\l(\frac{\vert \mu_{k}\vert^2}{W(k)}\r)
\eeq
which is expression~(\ref{eq:ps}) we have quoted in the text.

\end{document}